\documentclass{article}

\usepackage{graphicx, pdfpages}
\usepackage{latexsym}
\usepackage{mathtools}
\usepackage{amsmath, amsthm, amssymb}

\newsavebox{\astrutbox}
\sbox{\astrutbox}{\rule[-5pt]{0pt}{20pt}}

\newcommand{\be}{\begin{equation}}
\newcommand{\ee}{\end{equation}} 
\newcommand{\lb}{\label}

\newcommand{\OL}{\overline}

\newcommand{\const}{({\rm const.})}

\newcommand{\ba}{{\bf a}}

\newcommand{\bk}{{\bf k}}

\newcommand{\br}{{\bf r}}
\newcommand{\bu}{{\bf u}}

\newcommand{\bx}{{\bf x}}

\newcommand{\bJ}{{\bf J}}

\newcommand{\wt}{\widetilde}

\newcommand{\bomega}{{\mbox{\boldmath $\omega$}}}

\newcommand{\grad}{{\mbox{\boldmath $\nabla$}}}
\newcommand{\bdot}{{\mbox{\boldmath $\cdot$}}}

\newcommand{\btimes}{{\mbox{\boldmath $\times$}}}
\newcommand{\bzed}{{\mbox{\boldmath $0$}}}
\newcommand{\boell}{{\mbox{\boldmath $\ell$}}}

\begin{document}

\baselineskip=18pt
\begin{center}
\begin{LARGE}
{\bf Scale locality and the inertial range in compressible turbulence}\\
\end{LARGE}

\bigskip
\bigskip

Hussein Aluie\\
{\it
Applied Mathematics and Plasma Physics (T-5) \& Center for Non-linear Studies,\\
Los Alamos National Laboratory, MS-B258 Los Alamos, NM 87545, USA}

\bigskip
\bigskip

\begin{abstract}
We use a coarse-graining approach to prove that inter-scale transfer of kinetic energy in compressible turbulence
is dominated by local interactions. Locality here means that  interactions between disparate scales
decay at least as fast as a power-law function of the scale-disparity ratio.
In particular, our results preclude transfer of kinetic energy from
large-scales directly to dissipation scales, such as into shocks, in the limit of high Reynolds number turbulence as is commonly believed.
The results hold in broad generality, at any Mach number, for any equation of state,
and without the requirement of homogeneity or isotropy. The assumptions we make in our proofs
on the scaling of velocity, pressure, and density structure functions are weak and enjoy compelling empirical support.  
Under a stronger assumption on pressure dilatation co-spectrum, we show that 
\emph{mean} kinetic and internal energy budgets statistically decouple beyond a transitional
``conversion'' range. Our analysis demonstrates the existence of an ensuing inertial scale-range
over which  mean SGS kinetic energy flux becomes constant, independent of scale.
Over this inertial range, mean kinetic energy cascades locally and in a conservative fashion,
despite not being an invariant.
\end{abstract}
\end{center}

\vspace{0.5cm}

~~~{\bf Key Words:} Compressible turbulence; Scale locality; Cascade; Inertial range
\clearpage

\section{Introduction\lb{sec:Introduction}}

This paper is the second in a series which investigates the physical nature
of compressible turbulence. In the first paper \cite{Aluie10a}, we laid a framework
to study the coupling of scales in such flows and to analyze transfer 
of kinetic energy between different scales. The aim here is to explore 
whether this transfer of energy takes place through a scale-local cascade 
process, similar to incompressible turbulence.
This is of central importance in the subject because
scale locality is necessary to warrant the concept of an inertial range and to
justify the existence of universal statistics of turbulent fluctuations.

The traditional Richardson-Kolmogorov-Onsager picture of incompressible turbulence 
makes the fundamental assumption of a scale-local cascade process in which modes
all of a comparable scale $\sim \ell$ (differing at most by some fixed ratio, typically of order $2$) 
participate predominantly in the transfer of energy across scale $\ell$. This also implies that energy 
transfer is primarily between modes at comparable scales, with a ratio of order $2$.
If, furthermore, the cascade steps are chaotic processes
then it is expected that any ``memory'' of large-scale particulars of the system, such as 
geometry and large-scale statistics, or the specifics of microscopic dissipation,
will be ``forgotten.'' This gives rise to an \emph{inertial scale-range}
over which turbulent fluctuations have universal statistics
and the flow evolves under its own internal dynamics
without \emph{direct} communication with the largest or smallest scales in the system.

Therefore, scale locality is crucial to justify the existence of an inertial range
and its universal statistics, and is necessary for the 
physical foundation of large-eddy simulation (LES) modelling of turbulence. It motivates the belief that
models of subscale terms in the equations for large-scales can be of general utility, 
independent of the particulars of turbulent flows under study.

\cite{Kraichnan59,Kraichnan66,Kraichnan71} was the first to demonstrate locality in 
incompressible Navier-Stokes turbulence using detailed closure calculations. 
He showed that interactions between widely separated scales, $\ell_1 \ll \ell_2$,
decay as a power-law of their ratio, $(\ell_1/\ell_2)^\alpha$, where $\alpha > 0$.
Later on, \cite{Eyink05} was able to prove locality rigorously from the equations of motion 
and under very mild assumptions, without any closure or statistical averaging.
More recently,  \cite{AluieEyink09} proved locality using Fourier analysis and in 
\cite{AluieEyink10}, they 
showed that it also holds in incompressible magnetohydrodynamic turbulence. Furthermore, there has been several 
recent studies by \cite*{Domaradzki07a,Domaradzki09,AluieEyink09,AluieEyink10,Domaradzki10} 
which support the aforementioned theoretical results from direct numerical simulation (DNS) data.

No similar results, either theoretical or empirical, exist for compressible turbulence.
The idea of a cascade itself is without physical basis since kinetic energy is not
a global invariant of the inviscid dynamics. The notion of an inertial cascade-range 
in such flows is, therefore, still tenuous and unsubstantiated.
In this paper, we shall prove rigorously under modest assumptions that
inter-scale transfer of kinetic energy is indeed local in scale. Under a stronger
assumption, we will further show that kinetic energy cascades conservatively 
despite not being an invariant.

The outline of this paper is as follows. 
In \S\,\ref{sec:Preliminaries} we present preliminary definitions and discussion. 
In the course of proving locality of the cascade, we shall first prove that the 
subgrid scale flux (defined below) is dominated by scale-local interactions.
This is done by expressing the flux in terms of increments in \S\,\ref{sec:SGSfluxincrements},
then proving that increments themselves are scale-local in \S\,\ref{sec:Localityofincrements},
then, finally, showing how this leads to locality of the flux in \S\,\ref{sec:LocalityofFlux}.
In \S\,\ref{sec:Localityofcascade}, we discuss the implications on locality of the cascade
itself. We show that beyond a transitional ``conversion'' range, an
inertial range emerges over which kinetic energy cascades conservatively in
a scale-local fashion. 
In \S\,\ref{sec:ValidityofassumptionsA} and \S\,\ref{sec:ValidityofassumptionsB}, we argue for the 
validity of our assumptions based on empirical evidence and physical arguments.
In \S\,\ref{sec:intermittency} and \S\,\ref{sec:Shocks} we discuss common misconceptions
regarding scale locality in the presence of intermittency and shocks.
We summarize our main results in \S\,\ref{sec:summary} and defer 
mathematical details to Appendices \ref{ap:incrementrelations} and \ref{ap:Localityproofs}.

\section{Preliminaries\lb{sec:Preliminaries}}

\subsection{Governing dynamics}
We prove locality of kinetic energy transfer by a direct analysis of the compressible Navier Stokes equations, 
without use of any closure approximation. The equations are those of continuity (\ref{continuity}), 
momentum (\ref{momentum}), and either internal energy (\ref{internal-energy}) or total energy (\ref{total-energy}):
\begin{eqnarray} 
&\partial_t \rho& + \partial_j(\rho u_j) = 0 \lb{continuity} \\
&\partial_t (\rho u_i)& + \partial_j(\rho u_i u_j) 
= -\partial_i P +   \mu\partial_j(\partial_j u_i +\frac{1}{3}\partial_m u_m \delta_{ij} )  + \rho F_i   \lb{momentum}\\
&\partial_t (\rho e)&+ \partial_j\left\{\rho e u_j -\mu(u_m\partial_m u_j - u_j\partial_m u_m)\right\}
= -P\partial_j u_j + \mu |\partial_j u_i |^2 + \frac{\mu}{3} |\partial_j u_j |^2 -\partial_j q_j  ~~~~~~~~\lb{internal-energy}\\
&\partial_t (\rho E)&+ \partial_j(\rho E u_j) 
= -\partial_j (P u_j) + \mu\partial_j \{ u_i[ (\partial_j u_i + \partial_i u_j) - \frac{2}{3} \partial_m u_m \delta_{ij}] \} -\partial_j q_j 
+\rho u_i F_i  \lb{total-energy}\\
\nonumber\end{eqnarray}
Here, $\bu$ is velocity, $\rho$ is density, $e$ is internal energy per unit mass, $E=|\bu|^2/2 + e$ is total energy per unit mass, 
 $P$ is pressure, 
$\mu$ is dynamic viscosity, ${\bf F}$ is an external acceleration field stirring the fluid, 
${\bf q} = -\kappa \grad T$ is the heat flux with a conduction coefficient $\kappa$ and temperature $T$. 
For convenience, we have assumed a zero bulk viscosity even though all our analysis applies to the more general case. 
We have also assumed that $\mu = \nu\rho$ is independent of $\bx$. 

\subsection{Coarse-grained equations\lb{sec:coarsegraining}}
Following \cite{Germano92} and \cite{Eyink05},  we presented in a previous paper, \cite{Aluie10a},
a scale-decomposition based on coarse-graining which satisfies an \emph{inviscid criterion},
{\it i.e.} it guarantees that viscous momentum diffusion and kinetic energy dissipation are negligible
at large-scales. The decomposition yields a scale-range $L\gg \ell \gg \ell_\mu$ over which kinetic energy 
is immune from viscous dissipation and external injection by stirring. Here, $L$ denotes ``integral scale''
and $\ell_\mu$ denotes dissipation scale. 

Using the coarse-graining approach,
we can resolve dynamics \emph{both in scale and in space}. 
We define a coarse-grained or (low-pass) filtered field in $d$-dimensions as  
\be
\OL \ba_\ell(\bx) = \int d^d\br~ G_\ell(\br) \ba(\bx+\br),
\lb{filtering}\ee
where $G(\br)$ is a smooth convolution kernel which decays sufficiently rapidly for large $r$, 
and is normalized, $\int d^d\br ~G(\br)=1$. Its dilation $G_\ell(\br)\equiv \ell^{-d} G(\br/\ell)$ 
has its main support in a ball of radius $\ell$. We also define a complimentary high-pass filter
by
\be  \ba^{'}_\ell(\bx) = \ba(\bx)-\OL\ba_\ell(\bx).
\lb{high-pass}\ee
In the rest of our paper, we shall take the liberty of dropping subscript  $\ell$ whenever there is no risk of 
ambiguity.

In \cite{Aluie10a}, we proved that viscous momentum diffusion and kinetic energy dissipation are negligible
at large-scales when a Favre (or density-weighted) decomposition is employed. A Favre 
filtered field is weighted by density as
\be \tilde{f}_\ell(\bx) \equiv \frac{\OL{\rho f}_\ell(\bx)}{\OL\rho_\ell(\bx)}. \lb{FavreDef}\ee
The resultant large-scale dynamics for continuity and momentum are, respectively,
\be
\partial_t \OL{\rho} + \partial_i(\OL\rho \tilde{u}_i) = 0.
\lb{Favrecontinuity}\ee
\begin{eqnarray} 
\partial_t \OL\rho \tilde{u}_i + \partial_j (\OL\rho \tilde{u}_i~\tilde{u}_j )
& = & -\partial_j\left(\OL\rho~\tilde\tau(u_i,u_j)\right) -\partial_i\OL{P} \nonumber\\ 
& + & \mu \partial_j \{ [ (\partial_j \OL{u}_i + \partial_i \OL{u}_j) - \frac{2}{3} \partial_m \OL{u}_m \delta_{ij}] \} + \OL{\rho} \wt{F}_i,
\lb{Favremomentum}\end{eqnarray}
where 
\be\OL\rho\tilde\tau(u_i,u_j)\equiv \OL\rho(\wt{u_iu_j} - \tilde{u}_i~ \tilde{u}_i)\lb{Favrestress}\ee
is the \emph{turbulent stress} from the eliminated scales $<\ell$.
It is also straightforward to derive a kinetic energy budget for the large-scale, which reads
\be
\partial_t \OL\rho\frac{|\tilde\bu|^2}{2} + \grad\bdot\bJ_\ell 
= -\Pi_\ell -\Lambda_\ell + \OL{P}_\ell\grad\bdot\OL\bu_\ell
-D_\ell
+\epsilon^{inj},
\lb{largeKE}\ee
where $\bJ_\ell(\bx)$ is space transport of large-scale kinetic energy, $\Pi_\ell(\bx)+\Lambda_\ell(\bx)$,
is the subgrid scale (SGS) kinetic energy flux to scales $<\ell$, $-\OL{P}\grad\bdot\OL\bu$ is large-scale
pressure dilatation, $D_\ell(\bx)$ is viscous dissipation acting
on scales $>\ell$, and $\epsilon^{inj}(\bx)$ is the energy injected due to external stirring. These
terms are defined as 
\begin{eqnarray}
&\Pi_\ell(\bx)& = ~  -\OL{\rho}~ \partial_j\tilde{u}_i  ~\tilde\tau(u_i,u_j) ~~ \lb{flux1}\\[0.3cm]
& \Lambda_\ell(\bx)& = ~  \frac{1}{\OL\rho}\partial_j\OL{P}~\OL\tau(\rho,u_j) ~~ \lb{flux2}\\[0.3cm]
&D_\ell(\bx)&  = \mu \left[ \partial_j\tilde{u}_i ~\partial_j\OL{u}_i + \frac{1}{3}\partial_i\tilde{u}_i~\partial_j\OL{u}_j \right] \\[0.3cm]
&J_j(\bx)&  =  \OL\rho\frac{|\tilde\bu|^2}{2}\tilde{u}_j + \OL{P}\OL{u}_j + \tilde{u}_i\OL\rho\tilde\tau(u_i,u_j) 
-\mu \tilde{u}_i\partial_j\OL{u}_i -\frac{\mu}{3} \tilde{u}_j\partial_i\OL{u}_i    \\[0.4cm]
&\epsilon^{inj}(\bx)& =  \tilde{u}_i ~\OL{\rho} \wt{F}_i\lb{injectiondef}\\
\nonumber \end{eqnarray}
where 
\be  \OL\tau_\ell(f,g) \equiv\OL{(fg)_\ell}-\OL{f}_\ell\OL{g}_\ell \lb{tau-def} \ee
in expression (\ref{flux2}) is the $2^{nd}$-order \emph{generalized central moment} of fields $f(\bx),g(\bx)$ (see \cite{Germano92}). 
Equations (\ref{Favrecontinuity})-(\ref{largeKE}) describe the dynamics at scales $>\ell$, for arbitrary $\ell$,
at every point $\bx$ and at every instant in time. They hold for each realization of the flow without any statistical 
averaging.

The SGS flux is comprised of deformation work, $\Pi_\ell$, and baropycnal work, $\Lambda_\ell$,
which we discussed in some detail in \cite{Aluie10a}.
These represent the only two processes capable of direct transfer of kinetic energy \emph{across} scales.
Pressure dilatation, $-\OL{P}_\ell\grad\bdot\OL\bu_\ell$, does not contain any modes at scales $<\ell$ 
(or a moderate multiple thereof), at least for a filter kernel $\hat{G}(\bk)$ compactly supported in Fourier space.
Therefore, pressure dilatation cannot participate in the inter-scale transfer of kinetic energy and only 
contributes to conversion of large-scale kinetic energy into internal energy. This observation is one of the key
ingredients to proving scale locality.

\section{SGS flux in terms of increments\lb{sec:SGSfluxincrements}}

As was realized in the pioneering work of \cite{Eyink05}, there are two facts crucial for
scale locality of the SGS flux across $\ell$. First is that SGS flux can be written 
in terms of \emph{increments},
\be \delta \ba(\bx;\br)=  \ba(\bx+\br)-\ba(\bx),
\ee
for separation distances $|\br|<\ell$ (or some moderate multiple of $\ell$)
and do not depend on the absolute field $\ba(\bx)$. 

The SGS flux terms 
can be expressed in terms of increments by noting that 
gradient fields and central moments are related to increments as\footnote{
Relation (\ref{triplemomentrelation}) is based on an unpublished exact expression
due to G. L. Eyink (see \cite{EyinkNotes}). We repeat the details in Appendix \ref{ap:incrementrelations}.},
\begin{eqnarray}
\grad\OL{f}_\ell &=& {\cal O}[\delta f(\ell)/\ell], \lb{gradincrementrelation}\\
f'_\ell &=& {\cal O} [\delta f(\ell)], \lb{highincrementrelation}\\
\OL\tau_\ell(f,g) &=& {\cal O}[\delta f(\ell)~\delta g(\ell)],\lb{tauincrementrelation}\\
\grad \OL\tau_\ell(f,g) &=& {\cal O}[\delta f(\ell)~\delta g(\ell)/\ell],\lb{gradtauincrementrelation}\\
\OL\tau_\ell(f,g,h) &=& {\cal O}[\delta f(\ell)~\delta g(\ell)~\delta h(\ell)],\lb{triplemomentrelation}
\end{eqnarray}
where the symbol ${\cal O}$ stands for ``same order-of-magnitude as'' and
\be  \OL\tau_\ell(f,g,h) \equiv\OL{(fgh)}_\ell
-\OL{f}_\ell\OL{\tau}_\ell(g,h) -\OL{g}_\ell\OL{\tau}_\ell(f,h) -\OL{h}_\ell\OL{\tau}_\ell(f,g) -\OL{f}_\ell\OL{g}_\ell\OL{h}_\ell
\lb{triple-tau-def} \ee
is the $3^{rd}$-order \emph{generalized central moment} of fields $f(\bx),g(\bx),h(\bx)$ (see \cite{Germano92}). 
There are
rigorous versions of relations (\ref{gradincrementrelation})-(\ref{triplemomentrelation}).
See \cite{Eyink05} and Appendix \ref{ap:incrementrelations} below for details. 

Using relations (\ref{gradincrementrelation}) and (\ref{tauincrementrelation}), we can 
express baropycnal work as
\be\Lambda_\ell = {\cal O} \left[\frac{\delta P(\ell)}{\ell} \frac{\delta \rho(\ell)}{\OL\rho} \delta u(\ell)\right].
\lb{Lambdaincrements}\ee

In order to express $\Pi_\ell$ in terms of increments, 
we write down the following identities which are straightforward to verify:
\begin{eqnarray}
\tilde\bu&=&\OL\bu + \OL\tau(\rho,\bu)/\OL\rho \lb{identity1}\\
\partial_j\tilde{u}_i &=&  \partial_j\OL{u}_i + {\OL\rho}^{-1}\partial_j\OL\tau(\rho,u_i) - {\OL\rho}^{-2}\OL\tau(\rho,u_i)\partial_j\OL\rho
\lb{identity2}\\
\tilde\tau(u_i,u_j) &=& \OL\tau(u_i,u_j) + {\OL\rho}^{-1}\OL\tau(\rho,u_i,u_j) - {\OL\rho}^{-2}\OL\tau(\rho,u_i)\OL\tau(\rho,u_j)\lb{identity3}
\end{eqnarray}
It then follows that we can express deformation work (\ref{flux1}) as 
\begin{eqnarray}
\Pi_\ell(\bx) 
= &-&\OL{\rho} \left[\partial_j\OL{u}_i + {\OL\rho}^{-1}\partial_j\OL\tau(\rho,u_i) - {\OL\rho}^{-2}\OL\tau(\rho,u_i)\partial_j\OL\rho\right] 
\nonumber\\[0.3cm]
&\times& \left[\OL\tau(u_i,u_j) + {\OL\rho}^{-1}\OL\tau(\rho,u_i,u_j) - {\OL\rho}^{-2}\OL\tau(\rho,u_i)\OL\tau(\rho,u_j) \right].
\lb{KEflux1}\end{eqnarray}
From (\ref{KEflux1}) and relations (\ref{gradincrementrelation}),(\ref{tauincrementrelation})-(\ref{triplemomentrelation}), 
we can express deformation work as
\begin{eqnarray}
\Pi_\ell = {\cal O}\bigg[&\OL\rho&
\left[ \frac{\delta u(\ell)}{\ell} + \frac{\delta \rho(\ell)}{\OL\rho}\frac{\delta u(\ell)}{\ell} + \frac{\delta \rho^2(\ell)}{\OL\rho^2~}\frac{\delta u(\ell)}{\ell}\right]\nonumber\\[0.3cm]
&\times& \left[ \delta u^2(\ell) + \frac{\delta \rho(\ell)}{\OL\rho} \delta u^2(\ell) + 	\frac{\delta \rho^2(\ell)}{\OL\rho^2} \delta u^2(\ell)			\right]\bigg].
\lb{Piincrements}\end{eqnarray}
Expressions (\ref{Lambdaincrements}) and (\ref{Piincrements}) are not heuristic  estimates,
but are based on rigorous versions of (\ref{gradincrementrelation})-(\ref{triplemomentrelation}), whose
details can be found in \cite{Eyink05} and in Appendix \ref{ap:incrementrelations}.

\section{Locality of increments\lb{sec:Localityofincrements}}
Since $\Pi_\ell$ and $\Lambda_\ell$ can be expressed in terms of velocity, pressure, and density increments, it thus 
becomes sufficient to show that these increments themselves are 
scale-local. To establish this, we need the second requirement crucial for locality ---that
scaling properties of structure functions of velocity, pressure, and density increments 
are constrained by:
\begin{eqnarray}
\| \delta\bu(\br) \|_p &\sim& u_{rms} A_p (r/L)^{\sigma^u_p}, \,\,\,\,\,\,\,\,\,\,\,\,\,\,\,\,\, 0<\sigma^u_p<1,\lb{scaling1}\\
\| \delta P(\br) \|_p &\sim& P_{rms} B_p (r/L)^{\sigma^P_p}, \,\,\,\,\,\,~~~~\,\,\,\,\,\,\,\,\,\,\,\, \sigma^P_p<1,\lb{scaling2}\\
\| \delta\rho(\br) \|_p &\le& \rho_{rms} C_p (r/L)^{\sigma^\rho_p}, \,\,\,\,\,\,\,\,\,\,\,\,\,\,\,\,\,\, 0<\sigma^\rho_p\lb{scaling3}
\end{eqnarray}
for some dimensionless constants $A_p$, $B_p$, and $C_p$. The root-mean-square of a field $f(\bx)$
is denoted by $f_{rms} \equiv\langle f^2\rangle^{1/2}$, where $\langle\dots\rangle$ is a space average, $\frac{1}{V}\int d\bx(\dots)$.
Here, an $L_p$-norm $\| \cdot\|_p= \langle|\cdot|^p\rangle^{1/p}$ is just the traditional
structure function $S_p=\langle|\cdot|^p\rangle$ raised to the $1/p$-th power.
The scaling conditions (\ref{scaling1})-(\ref{scaling3}) are
well-established empirically in incompressible turbulent flows over intermediate scales $L\gg r \gg \ell_\mu$.
They also have strong empirical support
from many independent astronomical and numerical studies of compressible turbulent flows,
which we discuss in detail in \S\,\ref{sec:ValidityofassumptionsA}.
Note that condition (\ref{scaling3}) on the scaling of density increments
is only an upper bound. It only stipulates that the intensity of density fluctuations decays at smaller scales,
which is a very mild requirement and is readily satisfied in incompressible or nearly-incompressible
flows.

Scaling conditions (\ref{scaling1})-(\ref{scaling3}) reflect a structure of fields
at intermediate scales.
The constraints $\sigma_p^u <1$ and $\sigma_p^P <1$ indicate that velocity and pressure fields
 should be ``rough enough'' in a mean sense. They are not satisfied, for example, in laminar flows.
These conditions are used to prove that contributions from the very large scales
$\Delta$ to the flux across $\ell\ll \Delta$ are negligible or, in other words, that the flux is 
\emph{infrared local}.

On the other hand, constraints $\sigma_p^u >0$ and $\sigma_p^\rho >0$ 
indicate that the velocity and density fields are ``smooth enough'' in a mean sense.
They are not satisfied, for example, if the fields are dominated by small-scale fluctuations
with a non-decaying spectrum. 
These conditions are used to prove that contributions from the very small scales
$\delta$ to the flux across $\ell\gg\delta$ are negligible or, in other words, that
the flux is \emph{ultraviolet local}.

Notice that, unlike for the velocity and pressure fields, we do not stipulate that the 
density field be ``rough enough.'' Contributions to the flux across scale $\ell$
from the largest density scales $L\gg \ell$ need not be negligible, yet the flux
can still be scale-local.The underlying physical reason
 is simple; an energy flux across scale $\ell$ at a point $\bx$ will depend on
 the mass in a ball of radius $\ell$ around $\bx$. Mass is proportional to average 
 density $\OL\rho_\ell(\bx)$ in the ball, which is dominated by large scales: 
 $\OL\rho_\ell(\bx)= {\cal O}[\OL\rho_L(\bx)]={\cal O}[\rho_{rms}]$.
Indeed, for incompressible turbulence,
the only density scale present is a $\bk=0$ Fourier mode, the largest possible,
and the scale-local SGS flux is directly proportional to this density
mode.

Furthermore, we do not require that the pressure field be ``smooth enough,''
even though we expect $\sigma_p^P>0$. This is because pressure only
appears as a large-scale pressure-gradient in (\ref{flux2}). For any filter kernel
$\hat{G}(\bk)$ which is compact in Fourier space, contributions
from wavenumbers $Q\gg \ell^{-1}$ will be exactly zero and ultraviolet locality of 
$\grad\OL{P}_\ell$ is guaranteed without any scaling assumptions.

Under conditions (\ref{scaling1})-(\ref{scaling3}), proving 
scale locality of increments becomes simple and follows directly from \cite{Eyink05}.
For example, the contribution to any increment $\delta f(\ell)$ from scales 
$\Delta\geq\ell$ is represented by $\delta \OL{f}_\Delta(\ell)$. Here, $f(\bx)$
can denote either velocity or pressure field. Since the low-pass filtered 
field $\OL{f}_\Delta(\bx)$ is smooth, its increment may be estimated by Taylor expansion, 
and (\ref{gradincrementrelation}), and (\ref{scaling1}) or (\ref{scaling2}), as
\be \|\delta \OL{f}_\Delta(\ell)\|_p \simeq \|\boell \bdot(\grad\OL{f}_\Delta)\|_p
                                                =O\left[ \ell \,\, \frac{1}{\Delta} \left(\frac{\Delta}{L}\right)^{\sigma^f_p}\right]
                                                =O \left[\left(\frac{\ell}{L}\right)^{\sigma^f_p} \left(\frac{\ell}{\Delta}\right)^{1-\sigma^f_p}\right], \lb{infraredbound}\ee
and this is negligible for $\Delta\gg\ell$ as long as $\sigma^f_p<1.$ 
The notation $O(\dots)$ denotes a big-$O$ upper bound.

On the other hand,
the contribution to any increment $\delta f(\ell)$ from scales $\delta\leq\ell$ is represented 
by $\delta f_\delta'(\ell)$. Here, $f(\bx)$ can denote either velocity or density field.
 Since $f'_\delta = {\cal O}[f(x+\delta)-f(x)]$ from (\ref{highincrementrelation}), 
 scaling conditions (\ref{scaling1}) and (\ref{scaling3}) imply that
\be \|\delta f_\delta'(\ell)\|_p = O\left[\left(\frac{\delta}{L}\right)^{\sigma^f_p}\right]
                                        = O\left[\left(\frac{\ell}{L}\right)^{\sigma^f_p} \left(\frac{\delta}{\ell}\right)^{\sigma^f_p}\right], \lb{ultravioletbound}\ee
and this is negligible for $\delta\ll\ell$ as long as $\sigma^f_p>0$. 
For more details and for the careful proofs of these statements, see \cite{Eyink05}.

\section{Locality of flux\lb{sec:LocalityofFlux}}

Based on results in \S\,\ref{sec:SGSfluxincrements} and \S\,\ref{sec:Localityofincrements},
proving scale locality of the flux terms $\Pi_\ell$ and $\Lambda_\ell$ is straightforward.
To illustrate, consider flux due to baropycnal work, $\Lambda_\ell$ in (\ref{flux2}). 
It is a quartic quantity which depends on two density modes, one pressure mode, and one 
velocity mode. This dependence can be made more explicit by writing
$$\Lambda_\ell(\rho,P,\rho,\bu) \equiv \frac{1}{\OL\rho_\ell}\grad\OL{P}_\ell \bdot \OL\tau_\ell(\rho,\bu),
$$
where the first density argument $\Lambda(\rho,.,.,.)$ corresponds to the factor $1/\OL\rho$.

To prove ultraviolet locality of $\Lambda_\ell$, we need to show that contributions
from each of the four arguments $(\rho,P,\rho,\bu)$ at scales $\delta\ll\ell$
is negligible. 
It is obvious that $\OL\rho_\ell$ will have vanishing contribution 
(decaying faster than any power, or exactly zero for filter kernels $\hat{G}(\bk)$ compact in Fourier space) 
from scales $\delta\ll\ell$. It follows that its reciprocal $1/\OL\rho_\ell$ also has 
vanishing\footnote{
For any positive smooth filter kernel, the field $\OL\rho_\ell(\bx)$ is real analytic and non-zero for all $\bx$.
A well-known result from real analysis (see for e.g. \cite{KrantzParks02}) states that the reciprocal 
$1/\OL\rho_\ell$ is also real analytic and, therefore, its Fourier mode amplitudes 
decay faster than any power $n$ of wavenumber $k^{-n}$ as $k\to\infty$.}
contribution from very small scales $\delta\ll\ell$.
It is also obvious that $\grad\OL{P}_\ell$ will have vanishing contribution
from scales $\delta \ll\ell$.

What remains to be shown is that the last two arguments $( ., .,\rho,\bu)$
have negligible ultraviolet contributions. If we replace them by $\rho'_\delta$
and $\bu'_\delta$, respectively, we get from using H\"older's inequality that
\begin{eqnarray}
\| \Lambda_\ell(\rho,P,\rho'_\delta,\bu'_\delta) \|_p
&\le& \const \|\frac{1}{\OL\rho_\ell}\|_\infty  \|\grad\OL{P}_\ell\|_{3p}\|\delta \rho'_\delta(\ell)\|_{3p} \|\delta u'_\delta(\ell)\|_{3p}\nonumber\\
&=& O\left[ \left(\frac{\ell}{L}\right)^{\sigma^P_{3p}+\sigma^\rho_{3p}+\sigma^u_{3p}-1}\left(\frac{\delta}{\ell}\right)^{\sigma^\rho_{3p}+\sigma^u_{3p}}\right],
\lb{UVbound}\end{eqnarray}
which vanishes as $\delta/\ell\to 0$ for any $L_p$-norm  if $\sigma^\rho_{3p}+\sigma^u_{3p} > 0$. 
We used relation (\ref{tauincrementrelation}) to get 
$\|\OL\tau_\ell(\rho'_\delta,\bu'_\delta)\|_{3p/2}\le\const\|\delta \rho'_\delta(\ell)\|_{3p} \|\delta u'_\delta(\ell)\|_{3p}$
and result (\ref{ultravioletbound}) to arrive at the upper bound. 
A slight complication in our result is the additional overall factor $(\ell/L)^{\alpha}$, 
where $\alpha \equiv \sigma^{P}_{3p}+\sigma^{\rho}_{3p}+\sigma^{u}_{3p}-1$.
This can grow with decreasing $\ell$ if $\alpha < 0$, causing our
upper bound (\ref{UVbound}) to deteriorate at small scales. Such a
growth, however, can be easily offset by taking $\delta$ small enough: 
$\delta < \delta_*(\ell)=\ell\,(\ell/L)^{(1-\sigma^P_{3p}-\sigma^\rho_{3p}-\sigma^u_{3p})/(\sigma^\rho_{3p}+\sigma^u_{3p})}$.
We finally conclude that flux term $\Lambda_\ell$ is ultraviolet local
under scaling conditions (\ref{scaling1}),(\ref{scaling3}) for the velocity and density fields.

Next we prove infrared locality of $\Lambda_\ell(\rho,P,\rho,\bu)$ by replacing the 
pressure and velocity arguments with $\OL{P}_\Delta$ and $\OL\bu_\Delta$, respectively,
for $\Delta\gg\ell$. Using H\"older's inequality, we get
\begin{eqnarray}
\| \Lambda_\ell(\rho,\OL{P}_\Delta,\rho,\OL\bu_\Delta) \|_p
&\le& \const \|\frac{1}{\OL\rho_\ell}\|_\infty  \|\delta \OL{P}_\Delta(\ell)/\ell\|_{3p}\|\delta \rho(\ell)\|_{3p} \|\delta \OL{u}_\Delta(\ell)\|_{3p}\nonumber\\
&=&O\left[ \left(\frac{\ell}{L}\right)^{\sigma^P_{3p}+\sigma^\rho_{3p}+\sigma^u_{3p}-1}\left(\frac{\ell}{\Delta}\right)^{2-\sigma^P_{3p}-\sigma^u_{3p}}\right],
\lb{IRbound}\end{eqnarray}
which vanishes as $\ell/\Delta\to 0$ for any $L_p$-norm  if $\sigma^P_{3p}+\sigma^u_{3p} < 2$. 
We used relations (\ref{gradincrementrelation}) and (\ref{tauincrementrelation}) to get the inequality
and result (\ref{infraredbound}) to arrive at the upper bound. 
As in the ultraviolet case, there is a factor $(\ell/L)^{\alpha}$ which can 
cause our upper bound (\ref{IRbound}) to increase at small $\ell$ if $\alpha <0$.
Again, such an increase can be easily compensated by taking $\Delta$ large enough: 
$\Delta > \Delta_*(\ell)=\ell\,(\ell/L)^{(\sigma^P_{3p}+\sigma^\rho_{3p}+\sigma^u_{3p}-1)/(2-\sigma^P_{3p}-\sigma^u_{3p})}$.
We finally conclude that flux term $\Lambda_\ell$ is infrared local
under scaling conditions (\ref{scaling1}),(\ref{scaling2}) for the velocity and pressure fields.

Similarly, we can derive rigorous upper bounds on non-local contributions for each of the 9 terms in (\ref{KEflux1}) to 
prove scale locality of flux due to deformation work, $\Pi_\ell$.

\section{Locality of the cascade\lb{sec:Localityofcascade}}

In proving locality of the SGS flux, $\Pi_\ell+\Lambda_\ell$, 
 we did not need to make any assumptions about an equation of state for the fluid.
We also did not analyze the internal energy budget (\ref{internal-energy}). It is not obvious to us
how to best define a notion of scale for internal energy consistent with our scale-decomposition
of the flow field in \cite{Aluie10a}. Despite this shortcoming, we were still able to derive important 
results concerning locality.

Circumventing the aforementioned shortcoming was possible due to two facts.
First, we proved rigorously in \cite{Aluie10a} that viscous dissipation is negligible
at large-scales, which implies that large-scale kinetic energy 
does not couple to internal energy via viscous dynamics.
The only coupling that exists is via large-scale pressure dilatation. 
The second fact which aided us in proving locality
of the SGS flux is that large-scale pressure dilatation, 
$-\OL{P}_\ell\grad\bdot\OL\bu_\ell$, does not involve scales $<\ell$
and, therefore, it cannot transfer kinetic energy \emph{directly across} scales. 
Hence, pressure dilatation does not contribute to the SGS kinetic energy flux.

In principle, one could conjure a scenario in which mean large-scale
kinetic energy is converted to internal energy at scale $\sim\ell_1$ via 
$PD(\ell)\equiv-\langle\OL{P}_\ell\grad\bdot\OL\bu_\ell\rangle$ and 
is subsequently 
re-converted back, \emph{indirectly}, into mean kinetic energy at a much smaller scale $\sim\ell_2\ll\ell_1$.
In other words, as we continuously probe smaller scales $\ell$,
the function $PD(\ell)$ is positive at $\sim\ell_1$, then decreases and becomes zero at $\sim\ell_2$. 
The problem arises if such a process repeats itself, whereby $PD(\ell)$ oscillates indefinitely
with a non-decaying amplitude, as $\ell\to 0$.

As unlikely as such a scenario may appear, we do not know of a rigorous argument to disprove it under 
the weak assumptions (\ref{scaling1})-(\ref{scaling3}) we have already made.
It, therefore, precludes us from claiming at this point in the paper that our proof of a scale-local SGS 
flux implies rigorously
a scale-local cascade process in compressible turbulence. It is possible, however, to 
infer rigorously that the cascade is scale-local if we make one additional assumption which is,
albeit reasonable, not as weak as scaling conditions (\ref{scaling1})-(\ref{scaling3}).

\subsection{Pressure dilatation co-spectrum\lb{sec:Pressuredilatationcospectrum}}
The assumption we need to prove a scale-local cascade concerns the pressure dilatation co-spectrum, defined  as
\be E^{PD}(k)\equiv \sum_{k-0.5<|\bk|<k+0.5} -\hat{P}(\bk)\widehat{\grad\bdot\bu}(-\bk),
\lb{PDcospectrumdef}\ee
which we require to decay at a fast enough rate, 
\be |E^{PD}(k)| \le C\, u_{rms}\,P_{rms} \,(kL)^{-\beta},   \,\,\,\,\,\,\,\,\,\,\,\,\,\,\,\, \beta > 1.
\lb{scaling4}\ee
Here, $C$ is a dimensionless constant and $L$ is an integral scale.

In the limit of infinite Reynolds number, assumption (\ref{scaling4}) implies that
\emph{mean} pressure dilatation, $PD(\ell)$, asymptotes to a finite constant, 
$\theta\equiv-\langle P \grad\bdot\bu\rangle$, as $\ell\to 0$.
In other words, mean pressure dilatation, $PD(\ell)$, acting at scales $>\ell$ converges and becomes
independent of $\ell$ at small enough scales:
\be\lim_{\ell\to 0}PD(\ell) = \lim_{K \to \infty} \sum_{k<K} E^{PD}(k) = \theta,
\lb{PDresult}\ee
for wavenumber $K\approx \ell^{-1}$. In \S\,\ref{sec:ValidityofassumptionsB}, we shall give a physical
argument on why we expect (\ref{PDresult}) to hold.

We remark that condition (\ref{scaling4}) is sufficient but not necessary
for the convergence of $PD(\ell)$ in the limit of $\ell\to 0$. It is possible
for $E^{PD}(k)$, which is not sign-definite, to oscillate around $0$ as 
a function of $k$, such that the series $\sum_{k<K} E^{PD}(k)$
converges with $K \to \infty$ at a rate faster than what is implied by assumption (\ref{scaling4}).

\subsection{Conservative kinetic energy cascade\lb{sec:Conservativecascade}}
Saturation of \emph{mean} pressure dilatation, as expressed in (\ref{PDresult}), reveals 
that its role is to exchange \emph{large-scale} mean kinetic and internal energy
over a transitional ``conversion'' scale-range. At smaller scales beyond the conversion
range, mean kinetic and internal energy budgets statistically decouple.  
In other words, taking
$\ell_\mu\to 0$ first, then $\ell\to 0$,
the steady-state mean kinetic energy budget becomes,
\be \langle\Pi_\ell+\Lambda_\ell \rangle  = \langle\epsilon^{inj}\rangle - \theta.
\lb{inertialKEbudget}\ee
We stress that such a decoupling is statistical and does not imply that small scales evolve
according to incompressible dynamics. However, while small-scale compression and rarefaction can still 
take place pointwise, they yield a vanishing contribution to the \emph{space-average}. 

We denote the largest scale at which such statistical decoupling occurs by $\ell_c$. It may
be defined, for instance, as the scale at which $PD(\ell_c) = 0.95\,\theta$. Alternatively,
it may be defined as
\be \ell_c \equiv \frac{\sum_{\bk} k^{-1} E^{PD}(\bk)}{\sum_{\bk} E^{PD}(\bk)}.
\lb{PressDilscale}\ee
Over the ensuing scale-range, $\ell_c>\ell\gg\ell_\mu$, net pressure dilatation does 
not play a role, and if, furthermore, $\langle\epsilon^{inj}\rangle$ in (\ref{inertialKEbudget})
is localized to the largest scales as shown in \cite{Aluie10a}, then
$\langle \Pi_\ell+\Lambda_\ell\rangle$ will be a constant, independent of scale $\ell$.

A constant SGS flux implies that mean kinetic energy cascades conservatively to smaller scales,
despite not being an invariant of the governing dynamics. This is one of the major conclusions
of our paper. 
In particular, the scenario discussed in the two paragraphs preceding
\S\,\ref{sec:Pressuredilatationcospectrum} cannot ocure over $\ell_c>\ell\gg\ell_\mu$,
and kinetic energy  can only reach dissipation scales via the SGS flux, $\Pi_\ell+\Lambda_\ell$, through
a scale-local cascade process. We are therefore justified in calling scale-range $\ell_c>\ell\gg\ell_\mu$ the
inertial range of compressible turbulence.

\section{Discussion}
In proving locality of the SGS flux, $\Pi_\ell+\Lambda_\ell$, we did not invoke 
assumptions of homogeneity or isotropy. We only assumed the weak scaling
conditions (\ref{scaling1})-(\ref{scaling3}) on structure functions of velocity, pressure,
and density. The results also  apply to individual realizations of the flow, without the need
for ensemble averaging.

\subsection{Validity of assumptions (\ref{scaling1})-(\ref{scaling3})\lb{sec:ValidityofassumptionsA}}

We have proved through an exact analysis of the fluid equations
that scaling assumptions (\ref{scaling1})-(\ref{scaling3})
are sufficient to guarantee scale locality of SGS flux.
The ultimate source of these scaling properties is empirical evidence from experiments,
astronomical observations, and numerical simulations \footnote{ 
Our scaling conditions (\ref{scaling1})-(\ref{scaling3}) 
do not distinguish between
compressive $\bu^c$ and solenoidal $\bu^s$ components of the velocity field.
Whereas the contribution to structure functions in the incompressible limit will be predominantly 
from $\bu^s$, contributions from $\bu^c$ may be significant in general, as in the case of Burger's turbulence.}.

For incompressible turbulence, which may be viewed as a limiting case of our analysis,
the scaling of velocity and pressure structure functions (\ref{scaling1}),(\ref{scaling2}) has been 
well-established by a variety of independent studies  such as those by
\cite{Anselmetetal84,Benzietal95,Chenetal97,Sreenivasanetal96,HillBoratav97,GotohFukayama01,Xuetal07}.
Assumption (\ref{scaling3}) on density structure functions is trivially satisfied for a uniform density field.

As for compressible turbulence, the available data is also in compelling support of
our assumptions. Astronomical observations by  \cite{Armstrongetal81,Armstrongetal95}
 of radio wave scintillation in the interstellar medium,
characterized by highly supersonic turbulent flows, possibly up to Mach 20 (see for example \cite{Passotetal88}),
show that $2$nd-order density structure function scales with 
$\sigma_2^\rho \doteq 0.3$ over 5 decades in scale.
\cite{Stutzkietal98,Benschetal01} used velocity integrated spectral line maps
of several molecular clouds and found evident power-law scaling for the density with exponent
$0.3\le\sigma_2^\rho\le 0.4$. Much effort has also been expended to measure 
scaling of $2$nd-order velocity structure functions. Using spectroscopic surveys of molecular clouds, 
which give line-of-sight velocities from emission lines,
several independent studies by \cite*{Falgaroneetal92,MieschBally94,BruntHeyer02}; \cite{Padoanetal06}
found scaling exponents $\sigma_2^u \doteq 0.4$. More recently, \cite{HilyBlantetal08} 
measured scaling of structure functions up to $6$th-order and found 
$\sigma_1^u\doteq 0.54$, $\sigma_2^u\doteq  0.51$, $\sigma_3^u\doteq 0.49$,
$\sigma_4^u\doteq 0.47$, $\sigma_5^u\doteq 0.46$, $\sigma_6^u\doteq 0.45$.
Analysis of solar wind data has also yielded $0<\sigma_p^u<1$, for $1\le p\le 6$ 
(see for example \cite*{Podestaetal07} and \cite{Salemetal09}).

Alongside observational evidence, 
\cite{Kritsuketal07,Schmidtetal08, Schmidtetal09,Federrathetal10,PriceFederrath10}
carried out  several independent numerical
studies of forced compressible turbulence and calculated scaling of structure 
functions.
Their simulations employed an 
isothermal equation of state and reached a range of high turbulent Mach numbers
(based on $u_{rms}$ of velocity fluctuations), $M_t = 2.5 - 10$. 
They report power-law scaling exponents well within our required constraints,
$0<\sigma_p^\rho$ and $0<\sigma_p^u<1$ for $1\le p \le 6$.
One caveat of such simulations is that they do not resolve
viscous dynamics explicitly but rely on numerical schemes to deal with shocks and 
dissipation. Based on such considerations, they may not be deemed direct
numerical simulations but rather implicit LES with an uncontrolled subgrid model and, 
therefore, are not as reliable. DNS of compressible turbulence
cannot achieve simultaneously high Reynolds and Mach numbers due to resolution
limitations. 
The largest DNS to date we are aware of is that by \cite{PetersenLivescu10}.
It was on a $1024^3$ grid, had a Taylor Reynolds number $Re_\lambda=300$ 
and a turbulent Mach number  $M_t=0.3$ ---still in the subsonic regime.

We can also gain useful insights into the weakness of conditions (\ref{scaling1})-(\ref{scaling3})
through exact mathematical considerations.
From the definition of a structure function, $S_p^f(\ell)=\langle |\delta f(\ell)|^p\rangle$, for any field
$f(\bx)$ and its scaling
exponent, $\zeta_p^f = \liminf_{\ell\to 0}[\ln S_p^f(\ell)/\ln(\ell/L)]$ (see \cite{Eyink05,Eyink95,Eyink95a}),
it is known that $\zeta_p^f$ is a concave function of $p\in[0,\infty)$ (see for example \cite{Frisch,EyinkNotes}
for details). 
It follows that our scaling exponents $\sigma_p^f = \zeta_p^f /p$ are non-increasing
functions of $p$ (see \cite{EyinkNotes}). For example, if we have
$\sigma_1< 1$, then we are guaranteed $\sigma_p < 1$ for any $p\ge 1$. Similarly,
if $\sigma_q > 0$ for some $q>1$, then $\sigma_p > 0$ for any $p\le q$. 
Another known result states that if the $p$th-order moment of $f(\bx)$ exists, $\langle |f|^p\rangle<\infty$ for $p\ge 1$,
then the $p$th-order scaling exponent is non-negative, $\sigma_p^f\ge 0$ (see \cite{Frisch,EyinkNotes}).

To further put our scaling assumptions into perspective, 
a $2$nd-order structure function $S^u_2(\ell)\sim\ell^{2\sigma_2^u}$ is related to the spectrum $E^u(k)\sim k^{-n}$
with $n=2\sigma^u_2+1$. Therefore, condition $\sigma^u_2>0$  is equivalent to 
a constraint on the spectral exponent $n>1$. This condition on $n$ is the same as that required for 
a stationary velocity field in a bounded domain to have 
finite variance, $\langle|\bu|^2\rangle < \infty$, and a
power-law spectrum $k^{-n}$ for $k\in[k_0,\infty)$ in the limit $Re\to\infty$ (see \cite{Frisch}).

\subsection{Validity of assumption (\ref{scaling4})\lb{sec:ValidityofassumptionsB}}
The existence of a scale-local cascade over an inertial range is the main conclusion of 
this paper. However, to reach our result, we made the important assumption (\ref{scaling4})
which deserves more careful examination. 
Needless to say, the scaling of pressure dilatation co-spectrum is easily measurable from 
numerical simulations. The only reported measurement of this quantity we are aware of is
by \cite*{Leeetal06} shown in their figure 6(b). The authors had the same purpose in mind;
to check the scales at which pressure dilatation exchanges kinetic and internal energy. From their plot, they
concluded that such an exchange takes place only at the largest scales. While their conclusion
is in support of our postulate, their plot is on a log-linear scale which precludes
the inference of such a conclusion. It is possible for the co-spectrum to scale with $\beta \le 1$
in (\ref{scaling4}) while seeming to have most of its contribution from the largest scales.
The point we want to emphasize here is that it is $PD(\ell)$, the integral of the co-spectrum,
which needs converge and have most of its contribution from the largest scales.
We note that our condition (\ref{scaling4}) does not require a 
power-law scaling  ---only that $E^{PD}(k)$ decays at a rate faster than $\sim k^{-1}$.

It is not at all trivial why one should expect $PD(\ell) = -\langle\OL{P}_\ell\grad\bdot\OL\bu_\ell\rangle$ 
to converge at small scales. How can this be reconciled with the expectation that compression, 
as quantified by $\grad\bdot\bu$, would get more intense at smaller scales?
Indeed, it has been observed numerically that $(\grad\bdot\bu)_{rms}$ is an increasing
function of Reynolds number (see for example \cite*{Leeetal91}). The key point here is that our assumption (\ref{scaling4}) concerns 
\emph{spatially averaged} pressure dilatation. It is true that $\grad\bdot\bu(\bx)$, being a gradient,
derives most of its contribution from the smallest scales in the flow. However, since $P\grad\bdot\bu$
is not sign-definite, major cancellations can occur when space-averaging.
The situation is very similar to helicity co-spectrum in incompressible turbulence, 
$H(k)= \sum_{k-0.5<|\bk|<k+0.5} \hat{\bu}(\bk)\bdot\widehat{\grad\btimes\bu}(-\bk)$.
Here, the pointwise vorticity, $\bomega(\bx)=\grad\btimes\bu$, can also become unbounded in the limit
of infinite Reynolds number. However, numerical evidence shows that 
$\langle\bu\bdot\bomega\rangle$ remains finite with  Reynolds number 
and the co-spectrum $H(k)$ decays at a rate $\sim k^{-n}$,
with $n\approx 5/3 > 1$ (see for example \cite{Chenetal03} and \cite*{Kurienetal04}). 

We can offer a physical argument on why $PD(\ell)$ is expected to converge 
for $\ell\to 0$ as a result of cancellations from space-averaging.
The origin of such cancellations 
can be heuristically explained using decorrelation effects very similar to those studied
in \cite{EyinkAluie09} and \cite{AluieEyink09}. While the pressure in $\langle P\grad\bdot\bu\rangle$ derives
most of its contribution from the largest scales, $\grad\bdot\bu$ is dominated by the 
smallest scales in the flow. Therefore, pressure varies slowly in space, primarily at scales $\sim L$,
while $\grad\bdot\bu$ varies much more rapidly, primarily at scales $\ell_\mu\ll L$, leading to a 
decorrelation between the two factors. More precisely, the pressure $\OL{P}_\ell$ in 
$PD(\ell)$ may be approximated by $\OL{P}_\ell = {\cal O}[P_{rms}]= {\cal O}[\OL{P}_{L}]$ 
such that 
\begin{eqnarray}
\langle\OL{P}_\ell\grad\bdot\OL\bu_\ell\rangle 
\approx  \left\langle\OL{P}_L\grad\bdot\left(\OL{(\OL\bu_\ell)}_L +(\OL\bu_\ell)'_L \right)\right\rangle 
&\approx&  \langle\OL{P}_L\grad\bdot\OL\bu_L\rangle + \langle\OL{P}_L\rangle\langle\grad\bdot(\OL\bu_\ell)'_L\rangle. \nonumber
\end{eqnarray}
The first term in the last expression follows from $\OL{(\OL{\bu}_\ell)}_L \approx \OL\bu_L$, while
the second term is due to an approximate statistical independence between $\OL{P}_L$ and 
$\grad\bdot(\OL\bu_\ell)'_L \sim \delta u(\ell)/\ell$
which varies primarily at much smaller scales $\sim\ell \ll L$. If there is no transport beyond the 
domain boundaries or if the flow is either statistically homogeneous or isotropic, we get 
$\langle\grad\bdot(\OL\bu_\ell)'_L\rangle = 0$. The heuristic argument finally yields that pressure dilatation,
\be PD(\ell)=-\langle\OL{P}_\ell\grad\bdot\OL\bu_\ell\rangle \approx -\langle\OL{P}_L\grad\bdot\OL\bu_L\rangle,
\lb{PDresult2}\ee
becomes independent of $\ell$, for $\ell\ll L$. Expression (\ref{PDresult2}) corroborates our claim that the 
primary role of mean pressure dilatation is conversion of 
$\emph{large-scale}$ kinetic energy into internal energy and does not participate in the cascade dynamics
beyond a transitional ``conversion'' scale-range.

\subsection{A related study\lb{sec:RelatedStudy}}
An insightful and clever numerical study by \cite{Leeetal06}, which we mentioned in \S\,\ref{sec:ValidityofassumptionsB},
came to our attention at an advanced stage of writing this paper. The authors carried out DNS of
compressible isotropic turbulence at low Mach number decaying under the influence of a randomly
distributed temperature field. Some of the main conclusions are elegantly summarized by a schematic
in their figure 8. They assert that mean pressure dilatation only acts at the largest scales beyond which
mean kinetic energy cascades conservatively down to the viscous scales where it is dissipated into heat.
This conclusion is identical to the picture we arrived at in \S\,\ref{sec:Conservativecascade} and
\S\,\ref{sec:ValidityofassumptionsB}. However, their study does not address the issue of scale locality 
which forms a main theme of our paper.

On the other hand, their work goes beyond these statements and maintains that mean pressure dilatation couples
internal energy to irrotational (and not solenoidal) modes of the velocity field. They also observe that mean pressure dilatation 
is not sign-definite in time but tends to transfer energy from internal to kinetic energy after
time-averaging. Furthermore, they contend that the coupling of solenoidal and irrotational modes of the 
velocity field is weak and that each cascades separately to the viscous scales. They also
investigate the alignments between vorticity and gradients of pressure, density, and temperature.
All of these are essential issues which we do not tackle in our paper.

The paper of  \cite{Leeetal06} is a very valuable numerical investigation of the fundamental physics
of compressible turbulence. Yet, inspection of dissipation spectra in their figures 6(c,d) seem to indicate 
that the turbulent flows they studied were not fully developed. We believe that similar studies at higher Reynolds 
and Mach numbers, and under different controlled conditions are still needed to establish the 
findings of \cite{Leeetal06} as empirical facts.

\subsection{Intermittency, locality, and universality\lb{sec:intermittency}}

As we mentioned above,
scale locality is necessary to justify the notion of universality of 
intertial-range statistics. Kolmogorov's original 1941 theory 
of incompressible turbulence assumed statistical self-similarity of
inertial-range scales. Today, there is a general consensus based
on substantial empirical evidence that fluctuations in
turbulent flows are not statistically self-similar but are subject to 
intermittency corrections. For $p$-th order structure functions,
this is expressed as
\be S_p(\ell)\equiv \langle|\delta u(\ell)|^{p}\rangle \sim u_{rms}^{p} \left(\frac{\ell}{L}\right)^{\zeta_p}
=\left(\langle\epsilon\rangle \ell\right)^{p/3}  \left(\frac{\ell}{L}\right)^{\delta\zeta_p},
\lb{incompressibleintermittency}\ee
where average energy flux $\langle\epsilon\rangle$ (or dissipation) is empirically 
related to 
$u_{rms}$ and $L$ through the zeroth-law of incompressible turbulence, 
$\langle\epsilon\rangle = u_{rms}^3/L$, and we have
used $\zeta_p = p/3 + \delta\zeta_p$. Relation (\ref{incompressibleintermittency})
shows that for non-zero ``anomalous exponents'' $\delta\zeta_p$,
statistical averages at inertial-range scales $\ell\ll L$ are a function of integral length, $L$.
Colloquially, this means that inertial-range scales ``remember'' the number 
of ``cascade steps'', $\log_2(L/\ell)$, for energy in going from 
scale $L$ to scale $\ell$.

However, intermittency does not contradict scale locality or the existence of universal scaling.
It is known, for example, that the GOY shell model, in which scale 
interactions are local by construction, exhibits intermittency (see for example \cite{Biferale03}).
Despite remembering the number of cascade steps, there is no \emph{direct}
communication between inertial and integral length-scales due to scale locality.
It is precisely because individual cascade step are scale-local and depend
only upon inertial-range dynamics,
that it is possible to argue for 
universality of the scaling exponents $\zeta_p$, regardless of (and consistently with) 
intermittency
corrections. For a more detailed discussion of such issues, see \cite{EyinkNotes}.

\subsection{Shocks and locality\lb{sec:Shocks}}

An idea especially common in the astrophysical literature claims that in compressible turbulence
a ``finite portion'' of energy at a given scale must be dissipated directly into heat via shocks
rather than cascading in a local fashion (see for example \cite{McKeeOstriker07}).
Our analysis
shows that large-scale kinetic energy can only reach dissipation scales through 
SGS flux, $\Pi_\ell + \Lambda_\ell$, which
we have have proved to be scale-local provided the weak scaling conditions 
(\ref{scaling1})-(\ref{scaling3}) are satisfied. Therefore, in order for 
large-scale kinetic energy to dissipate into heat non-locally, it is necessary to violate
(\ref{scaling1}),(\ref{scaling3}) such that $\sigma^u_p\le0$ or
$\sigma^\rho_p\le0$ for $p\le7$ to break down ultraviolet locality.
Having $\sigma^{u,\rho}_p=0$ implies that the mean intensity of velocity or density
fluctuations does not decay at smaller scales. All empirical evidence discussed in
subsection \ref{sec:ValidityofassumptionsA} seems to rule out such a possibility.

The situation in compressible turbulence is similar to that of incompressible MHD turbulence 
where discontinuities in the magnetic field, i.e. current sheets, are pervasive. However, 
 \cite{AluieEyink10} proved rigorously, under scaling conditions analogous
to (\ref{scaling1})-(\ref{scaling3}), that the cascade is local in scale and, furthermore, provided
numerical support from high-resolution DNS. 

Another elucidating example is that of Burger's turbulence, in which viscous 
dissipation takes place only in shocks (which have zero volume when the Reynolds number is infinite)
and vanishes everywhere else. However, scaling
exponents of velocity increments are known to be $\sigma^u_p = 1/p$ for $p\ge 1$, 
which satisfy the condition $0<\sigma^u_p<1$ for any $1\le p<\infty$,
and the proof of locality applies to Burger's energy cascade as well. In fact, the same conclusion
was pointed out by \cite{Kraichnan74} where he 
stated that the cascade in turbulence can be scale-local despite the presence of 
coherent discontinuous structures. He specifically discussed Burger's turbulence 
and showed that the energy is transfered to smaller scales in a local cascade process.
As we mentioned in the introduction \S\,\ref{sec:Introduction}, Kraichnan at the time
had realized through his closure theory that scale locality depends on the exponent of the 
spectrum power-law and not on coherence properties.
Furthermore, scale locality is perfectly consistent with the possibility
that all dissipation takes place only in shocks and singular structures, 
over a set of zero volume in the limit of $\mu\to 0$.
It is well-known for this to be the case in Burger's flow. The point here being that scale locality is an 
inertial-range property of
SGS flux  which transfers energy across scales and not of
viscous dissipation which converts energy into heat.

\section{Summary\lb{sec:summary}}
We proved that in compressible turbulence,
the SGS flux responsible for direct transfer of kinetic energy across scales is dominated
by scale-local interactions. This was achieved through rigorous upper bounds on the non-local
contributions, and under weak assumptions (\ref{scaling1})-(\ref{scaling3}). No assumptions
of homogeneity or isotropy were invoked and the results hold for any equation of state of the fluid.

We also showed that, in the limit of high Reynolds number, scale locality of the SGS flux implies 
a scale-local cascade of kinetic energy if condition (\ref{scaling4}) on pressure dilatation co-spectrum is satisfied. 
In particular, condition (\ref{scaling4}) implies that beyond a transitional ``conversion'' scale-range,
there exists an inertial scale-range over which
mean kinetic and internal energy budgets statistically decouple 
and the mean SGS flux, $\langle\Pi_\ell +\Lambda_\ell\rangle$, becomes a constant, independent of scale $\ell$.
Our result in \S\,\ref{sec:Localityofcascade} demonstrates that kinetic energy 
cascades conservatively despite not being an invariant.  

We remark that the extent of the conversion range 
could be an increasing function of Mach number and/or the ratio of compressive-to-solenoidal kinetic energy.
If so, it would have an immediate bearing on the interpretation of results such as those in figure 2 of 
\cite{Padoanetal04}, where the Mach number is varied while maintaining a fixed Reynolds number.
In such studies, an increasing Mach number may lead to the conversion range eroding away the 
finite inertial range present in simulations. Measurements of power-law exponents over the conversion
range may not reflect an asymptotic scaling which would otherwise appear at sufficiently high Reynolds numbers.
The problem is that of an ordering of limits; the physically interesting order being one in which we take 
$Re\to \infty$ first, followed by $M_t \to \infty$.

In summary, we conclude that there exists an inertial range in high Reynolds number  compressible turbulence
over which kinetic energy reaches dissipation scales through a conservative and scale-local cascade process.
This precludes the possibility for transfer of kinetic energy from the large-scales directly to dissipation scales,
such as into shocks, at arbitrarily high Reynolds numbers as is commonly believed. 
Our work makes several assumptions and predictions which can be tested numerically. Our locality
results concerning the SGS flux can be verified in a manner very similar to what was done in \cite{AluieEyink09}
and \cite{AluieEyink10}. We also invite numerical tests of assumption (\ref{scaling4}) on
the scaling of pressure dilatation co-spectrum. 
Verifying (\ref{scaling4}) or (\ref{PDresult}) under a variety of controlled conditions would substantiate the 
idea of statistical decoupling between mean kinetic and internal energy budgets. This would have potentially 
significant implications on devising reduced models of compressible turbulence as well as providing physical 
insight into this rich problem.

\noindent {\small
{\bf Acknowledgements.} 
I am indebted to G. L. Eyink for invaluable discussions 
on turbulence over many years. I want to especially thank D. Livescu for his
input on compressible and variable-density turbulence. I also thank S. S. Girimaji
for bringing to my attention previous studies relevant to this work and X. Asay-Davis 
for useful comments on an earlier version of the manuscript. 
I am appreciative of
S. Kurien, S. Li, and H. Li for their encouragement and support during this project.
This research was performed under the auspices of the U.S. Department of
Energy at LANL under Contract No. DE-AC52-06NA25396 and supported
by the LANL/LDRD program.
}

\appendix

\section{Details of relations (\ref{gradincrementrelation})-(\ref{triplemomentrelation})\label{ap:incrementrelations}}

We repeat from \cite{Eyink05} and \cite{EyinkNotes}
the details of expressing gradients and central moments in terms of increments.

Relation (\ref{gradincrementrelation}) is a result of integration by parts and rewriting a large-scale gradient as:
\begin{eqnarray}
\grad \OL{f}_\ell (\bx) &=&-\frac{1}{\ell} \int d\br (\grad G)_\ell(\br)  f(\bx+\br) \nonumber\\
&=&-\frac{1}{\ell} \int d\br (\grad G)_\ell(\br)  [f(\bx+\br)  -f(\bx)], 
\lb{app:gradincrementrelation}\end{eqnarray}
where $(\grad G)_\ell(\br) = \ell^{-d} (\grad G)(\br/\ell)$ in $d$-dimensions and we have used condition
$\int d\br \grad G (\br) = \bzed$ to arrive at the last equality.

High-pass filtered fields in relation (\ref{highincrementrelation}) can be expressed in terms of increments as:
\begin{eqnarray}
 f'_\ell (\bx) &=& -\int d\br\, G_\ell(\br)  [f(\bx+\br)-f(\bx)] \nonumber\\
&=&-\langle\delta f(\bx;\br)\rangle_\ell, 
\lb{app:highincrementrelation}\end{eqnarray}
where $\langle\delta f(\bx;\br)\rangle_\ell \equiv \int d\br G_\ell(\br)  \delta f(\bx;\br)$ is an average over separations 
$\br$ in a ball of radius of order $\ell$ centered at $\bx$.

The $2^{nd}$-order central moment relation (\ref{tauincrementrelation}) is due to \cite*{Constantinetal94} and is 
straightforward to verify:
\begin{eqnarray}
\OL\tau_\ell\left(f \left(\bx \right),g \left(\bx\right) \right) 
= \langle\delta f(\bx;\br) \delta g(\bx;\br)\rangle_\ell -\langle\delta f(\bx;\br) \rangle_\ell \langle\delta g(\bx;\br)\rangle_\ell.
\lb{app:tauincrementrelation}\end{eqnarray}

Relation (\ref{gradtauincrementrelation}) follows from (\ref{app:tauincrementrelation})
through an integration by parts:
\begin{eqnarray}
\grad \OL\tau_\ell\left(f \left(\bx \right),g \left(\bx\right) \right) 
= -\frac{1}{\ell}\bigg\{
&&\int d\br (\grad G)_\ell(\br) \delta f(\bx;\br) \delta g(\bx;\br) \nonumber\\ 
&&-\int d\br (\grad G)_\ell(\br) \delta f(\bx;\br) \int d\br' G_\ell(\br') \delta g(\bx;\br')  \nonumber\\
&&- \int d\br G_\ell(\br) \delta f(\bx;\br) \int d\br' (\grad G)_\ell(\br') \delta g(\bx;\br')\bigg\}.
\lb{app:gradtauincrementrelation} \end{eqnarray}

The $3^{rd}$-order central moment in relation (\ref{triplemomentrelation}) is due to \cite{EyinkNotes} and is 
straightforward to verify:
\begin{eqnarray}
\OL{\tau}_\ell(f, g, h)(\bx)
&=& \langle\delta f(\bx;\br) \delta g(\bx;\br) \delta h(\bx;\br)\rangle_\ell   \nonumber\\
&-&\langle\delta f(\bx;\br) \rangle_\ell \langle\delta g(\bx;\br) \delta h(\bx;\br)\rangle_\ell  \nonumber\\
&-&\langle\delta g(\bx;\br) \rangle_\ell \langle\delta f(\bx;\br) \delta h(\bx;\br)\rangle_\ell  \nonumber\\
&-&\langle\delta h(\bx;\br) \rangle_\ell \langle\delta f(\bx;\br) \delta g(\bx;\br)\rangle_\ell  \nonumber\\
&+&2\langle\delta f(\bx;\br) \rangle_\ell \langle\delta g(\bx;\br) \rangle_\ell \langle\delta h(\bx;\br) \rangle_\ell.
\lb{app:triplemomentrelation}\end{eqnarray}

\section{Locality proofs\label{ap:Localityproofs}}

The proofs for locality follow closely those of \cite{Eyink05}. We will repeat briefly the ones which were explicitly
discussed in the latter work and also give some details on locality of new terms not present in the \cite{Eyink05} treatment.

The filter kernel $G(\br)$ used in the proofs is smooth and decays faster than any power $r^{-p}$ as $r\to\infty$.
It is also possible (but not required) to have both it and its Fourier transform $\hat{G}(\bk)$ be positive
and, furthermore, to have the latter compactly supported inside a ball of radius $1$ about the origin
in Fourier space.

\subsection{Ultraviolet locality}
Ultraviolet locality means that contributions to the energy flux across $\ell$ from scales
$\delta \ll \ell$ decay at least as fast as $\delta^\alpha$, for some $\alpha>0$.
We will now show that each of the factors in SGS flux terms (\ref{flux2}) and (\ref{KEflux1}) is ultraviolet local.

Non-local ultraviolet contributions to a large-scale gradient of field $f(\bx)$ can be shown to be 
bounded using (\ref{app:gradincrementrelation})-(\ref{app:highincrementrelation}) as was proved by \cite{Eyink05}:
\be
\| \grad\OL{f'_\delta} \|_p 
\le  \frac{2}{\ell} \int d\br\, |(\grad G)_\ell(\br)|~ \| f'_\delta(\bx)  \|_p
=O\left(\frac{\delta^{\sigma_p^f}}{\ell}\right).
\lb{AppendixUV1}\ee
Notation $O(\dots)$ denotes a big-$O$ upper bound.
In fact, $\| \grad\OL{\bu'_\delta} \|_p=0$ for a filter $\hat{G}(\bk)$ compactly supported in Fourier space.

Non-local ultraviolet contributions to a second-order central moment of fields $f(\bx)$ and $g(\bx)$
can be bounded using (\ref{app:highincrementrelation}),(\ref{app:tauincrementrelation}) as was proved by \cite{Eyink05}. For 
$1/p=1/r +1/s$, we have
\begin{eqnarray}
\| \OL{\tau}_\ell(f'_\delta, g'_\delta)\|_p  
\le & &   4 \int d\br |G_\ell(\br)| ~\| f'_\delta(\bx) \|_r  \| g'_\delta(\bx) \|_s \nonumber\\
&+& 4 \int d\br |G_\ell(\br)|~ \| f'_\delta(\bx) \|_r   \int d\br |G_\ell(\br)| ~ \| g'_\delta(\bx) \|_s\nonumber\\
= & & O\left( \delta^{\sigma_r^f + \sigma_s^g}  \right).
\lb{AppendixUV2}\end{eqnarray} 


Similarly, using (\ref{app:highincrementrelation}),(\ref{app:gradtauincrementrelation}) we can show that non-local 
ultraviolet contributions to the gradient of a second-order central moment  are bounded.
For $1/p=1/r +1/s$, we have
\begin{eqnarray}
\|  \grad\OL{\tau}_\ell(f'_\delta, g'_\delta)\|_p 
\le& &   \frac{4}{\ell} \bigg\{  \int d\br |(\grad G)_\ell(\br)| ~\| f'_\delta(\bx) \|_r  \| g'_\delta(\bx) \|_s\nonumber\\
&+&  \int d\br |(\grad G)_\ell(\br)|~ \| f'_\delta(\bx) \|_r   \int d\br |G_\ell(\br)| ~ \| g'_\delta(\bx) \|_s  \nonumber\\
&+&  \int d\br |G_\ell(\br)|~ \| f'_\delta(\bx) \|_r   \int d\br |(\grad G)_\ell(\br)| ~ \| g'_\delta(\bx) \|_s  \bigg\}\nonumber\\
= & & O\left( \frac{\delta^{\sigma_r^f + \sigma_s^g}}{\ell}  \right).
\lb{AppendixUV3}\end{eqnarray}


The non-local ultraviolet contributions to a third-order central moment of fields $f(\bx)$, $g(\bx)$, and $h(\bx)$
can be bounded using (\ref{app:highincrementrelation}),(\ref{app:triplemomentrelation}). For $1/p=1/r +1/s + 1/t$, we have
\begin{eqnarray}
\| \OL{\tau}_\ell(f'_\delta, g'_\delta, h'_\delta)\|_p  
&\le& 8 \int d\br |G_\ell(\br)| ~\| f'_\delta(\bx) \|_r  \| g'_\delta(\bx) \|_s \| h'_\delta(\bx) \|_t  \nonumber\\
&+& 8 \int d\br |G_\ell(\br)|~ \| f'_\delta(\bx) \|_r   \int d\br |G_\ell(\br)| ~ \| g'_\delta(\bx) \|_s \| h'_\delta(\bx) \|_t  \nonumber\\
&+& 8 \int d\br |G_\ell(\br)|~ \| g'_\delta(\bx) \|_r   \int d\br |G_\ell(\br)| ~ \| f'_\delta(\bx) \|_s \| h'_\delta(\bx) \|_t  \nonumber\\
&+& 8 \int d\br |G_\ell(\br)|~ \| h'_\delta(\bx) \|_r   \int d\br |G_\ell(\br)| ~ \| f'_\delta(\bx) \|_s \| g'_\delta(\bx) \|_t  \nonumber\\
&+& 16 \int d\br |G_\ell(\br)|~ \| f'_\delta(\bx) \|_r  \int d\br |G_\ell(\br)|~ \| g'_\delta(\bx) \|_s  \int d\br |G_\ell(\br)|~ \| h'_\delta(\bx) \|_t    \nonumber\\
&=& O\left( \delta^{\sigma_r^f + \sigma_s^g + \sigma_t^h}  \right).
\lb{AppendixUV4}\end{eqnarray} 


\subsection{Infrared locality}
Infrared locality means that contributions to the energy flux across $\ell$ from scales
$\Delta \gg \ell$ decay at least as fast as $\Delta^{-\alpha}$, for some $\alpha>0$.
We will now show that each of the factors in SGS flux terms (\ref{flux2}) and (\ref{KEflux1}), 
except for the density, is infrared local. 
As we have discussed above, it is physically expected that the kinetic energy cascade
will depend on density  fluctuations at the largest scales.

Non-local infrared contributions to a large-scale gradient of field $f(\bx)$ can be 
bounded using (\ref{app:gradincrementrelation}) as was shown by \cite{Eyink05}:
\be
\| \OL{\left(\grad\OL{f}_\Delta \right)_\ell} \|_p 
\le  \int d\br' \, |G_\ell(\br')| ~\frac{1}{\Delta}\int d\br\, |(\grad G)_\Delta(\br)|~ \| \delta f(\bx;\br)  \|_p
=O\left(\Delta^{\sigma_p^f-1}\right).
\lb{AppendixIR1}\ee
Non-local infrared contributions to a second-order central moment of fields $f(\bx)$ and $g(\bx)$
can be bounded using (\ref{app:tauincrementrelation}) as was shown by \cite{Eyink05}. For 
$1/p=1/r +1/s$, we have
\begin{eqnarray}
\| \OL{\tau}_\ell(\OL{f}_\Delta, \OL{g}_\Delta)\|_p 
\le  & & \int d\br |G_\ell(\br)| ~\| \delta \OL{f}_\Delta(\bx;\br) \|_r  \| \delta \OL{g}_\Delta(\bx;\br) \|_s  \nonumber\\
&+&  \int d\br |G_\ell(\br)|~ \| \delta \OL{f}_\Delta(\bx;\br) \|_r   \int d\br |G_\ell(\br)| ~ \| \delta \OL{g}_\Delta(\bx;\br) \|_s \nonumber\\
= & & O\left( \Delta^{\sigma_r^f + \sigma_s^g-2}~ \ell^2 \right),
\lb{AppendixIR2}\end{eqnarray} 
where the last step follows from $\| \delta \OL{g}_\Delta(\bx;\br) \|_s = O(r\Delta^{\sigma_s^g-1})$ (see \cite{Eyink05}).

Using (\ref{app:gradtauincrementrelation}), we can also show that non-local 
infrared contributions to the gradient of a second-order central moment  are bounded.
Such a term only appears in the form $\grad\bdot\OL{\tau}_\ell(\rho,\bu)$ in (\ref{KEflux1}).
For $1/p=1/r +1/s$, we have
\begin{eqnarray}
\|  \grad\bdot\OL{\tau}_\ell(\rho,\OL\bu_\Delta)\|_p 
\le & &  \frac{1}{\ell} \bigg\{  \int d\br |(\grad G)_\ell(\br)|\bdot ~\| \delta \rho(\bx;\br) \|_r  \| \delta \OL{\bu}_\Delta(\bx;\br)\|_s\nonumber\\
&+&  \int d\br |(\grad G)_\ell(\br)|~ \| \delta \rho(\bx;\br) \|_r  \bdot \int d\br |G_\ell(\br)| ~ \| \delta \OL{\bu}_\Delta(\bx;\br) \|_s  \nonumber\\
&+&  \int d\br |G_\ell(\br)|~ \| \delta \rho(\bx;\br) \|_r   \int d\br |(\grad G)_\ell(\br)|\bdot  \| \delta \OL{\bu}_\Delta(\bx;\br) \|_s  \bigg\}\nonumber\\
= & & O\left( \Delta^{\sigma_s^u -1}~\ell^{\sigma^\rho_r}  \right).
\lb{AppendixIR3}\end{eqnarray} 

We can also show infrared locality of a third-order central moment. Such a term appears in
deformation work (\ref{KEflux1}) in the form $\OL{\tau}_\ell(\rho,\bu,\bu)$.
Non-local infrared velocity contributions can be bounded using (\ref{app:triplemomentrelation}). 
For $1/p=1/r +1/s + 1/t$, we have
\begin{eqnarray}
\| \OL{\tau}_\ell(\rho, \OL{\bu}_\Delta, \OL{\bu}_\Delta)\|_p  
&\le&  \int d\br |G_\ell(\br)| ~\| \delta\rho(\bx;\br) \|_r  \| \delta\bu_\Delta(\bx;\br) \|_s \| \delta\bu_\Delta(\bx;\br) \|_t  \nonumber\\
&+&  \int d\br |G_\ell(\br)|~ \| \delta\rho(\bx;\br) \|_r   \int d\br |G_\ell(\br)| ~ \| \delta\bu_\Delta(\bx;\br) \|_s \| \delta\bu_\Delta(\bx;\br) \|_t  \nonumber\\
&+&  \int d\br |G_\ell(\br)|~ \| \delta\bu_\Delta(\bx;\br) \|_r   \int d\br |G_\ell(\br)| ~ \| \delta\rho(\bx;\br) \|_s \| \delta\bu_\Delta(\bx;\br) \|_t  \nonumber\\
&+&  \int d\br |G_\ell(\br)|~ \| \delta\bu_\Delta(\bx;\br) \|_r   \int d\br |G_\ell(\br)| ~ \| \delta\rho(\bx;\br) \|_s \| \delta\bu_\Delta(\bx;\br) \|_t  \nonumber\\
&+& 2 \int d\br |G_\ell(\br)|~ \| \delta\rho(\bx;\br) \|_r  \int d\br |G_\ell(\br)|~ \| \delta\bu_\Delta(\bx;\br) \|_s  \int d\br |G_\ell(\br)|~ \| \delta\bu_\Delta(\bx;\br) \|_t    \nonumber\\
&=& O\left( \Delta^{\sigma_s^u + \sigma_t^u -2} ~\ell^{\sigma^\rho_r+2}  \right).
\lb{AppendixUV4}\end{eqnarray}


\end{document}